\documentclass[onecolumn]{elsart}
\usepackage{graphicx}
\usepackage{dcolumn}
\usepackage{bm}
\usepackage{epsfig}
\input epsf
\epsfclipon

\newcommand{\be}{\begin{equation}}
\newcommand{\ee}{\end{equation}}
\newcommand{\bc}{\begin{center}}
\newcommand{\ec}{\end{center}}
\newcommand{\bi}{\begin{itemize}}
\newcommand{\ei}{\end{itemize}}
\newcommand{\ba}{\begin{eqnarray}}
\newcommand{\ea}{\end{eqnarray}}

\newcommand{\ignore}[1]{}
\newcommand{\mean}[1]{\left\langle #1 \right\rangle}

\begin{document}
\begin{frontmatter}
\title{Voter model on Sierpinski fractals}
\author{Krzysztof Suchecki and Janusz A. Ho{\l}yst\corauthref{holyst}}
\corauth[holyst]{Corresponding author. Tel.: +48 22 660 7133; fax: +48 22 628 2171}
\ead{jholyst@if.pw.edu.pl}
\address{Faculty of Physics and Center of Excellence
for Complex Systems Research, Warsaw University of Technology,
Koszykowa 75, PL-00-662 Warsaw, Poland}
\date{\today}
\begin{abstract}
We investigate the ordering of voter model on fractal lattices: Sierpinski Carpets and Sierpinski Gasket. We obtain a power law ordering, similar to the behavior of one-dimensional system, regardless of fractal ramification.
\end{abstract}
\maketitle
\end{frontmatter}

\section{Introduction}
The Ising model is a well known dynamical model that was investigated in complex networks and fractal structures \cite{intro1,intro2,mandelbrot,ising1,ising2}.
However, aside from that model, there are many other possible dynamics, sharing little in common with behavior of the Ising model.
The voter model is an example of such a model, that exhibits different qualities at a very basic level.
Unlike the Ising model, the voter model has no surface tension and defines a broad universality class \cite{universality}.
While the Ising model dynamics has been studied on fractal lattices \cite{mandelbrot,ising1,ising2} little is known about the behavior of voter model in such geometries.\\
We have investigated the behavior of the voter model on Sierpinski carpets and on Sierpinski gasket.
It is known \cite{krapivsky} that the evolution of the voter model depends on the dimensionality of the lattice. For a large time $t$ the ordering process obeys the following equations
\be
\label{krap}
\rho(t) \sim \left\{
\begin{array}{ll}
t^{-\alpha},& D<2\\
(\ln t)^{-1},& D=2\\
1,& D>2\\
\end{array} \right.
\ee
where $\rho$ is a fraction of links that form interfaces, i.e. they connect opposite spins, $D$ is dimensionality and $t$ is time.
The predicted exponent $\alpha=1-D/2$ was calculated for a lattice with {\em integer} dimensionality ($D=1$).\\
We will focus on the question, whether the dynamics on the fractal lattices is the same, or are there other rules governing them.

\section{Models}

\begin{figure}
 \centerline{\epsfig{file=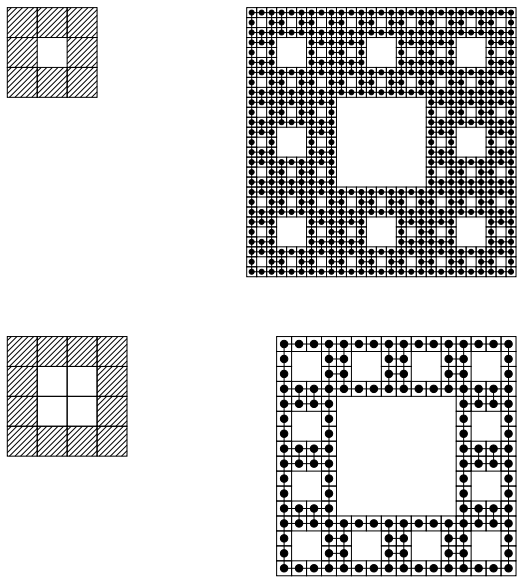,width=.6\columnwidth}}
    \caption{The construction of Sierpinski Carpet (SC) fractal network. Two different basic patterns are on the left, and corresponding fractal networks of level $3$ and $2$ are on the right. The hatched area of patterns are full positions.}
    \label{dywan}
\end{figure}

The voter model is a very simple model of opinion formation. Nodes in the network are agents, each one having an opinion.
There are only two possible opinions, and typically they are considered as +1 and -1, just as Ising spins.
The dynamic rule is simple --- the node opinion changes to an opinion of one randomly chosen neighbor.\\
The implementation is following: we choose one node at random, and then one of its neighbors randomly.
The first node assumes the state of the second. One time step of the dynamics corresponds to the number of individual node updates equal to the number of nodes in the network, so on average each node is updated once every time step.\\
We investigate the voter model behavior on two fractal networks: Sierpinski Carpet (SC) and Sierpinski Gasket (SG).
The SC is constructed according to a chosen basic pattern. The pattern is a square, divided into $n\times n$ squares that could be full or empty (Fig.\ref{dywan}).
First, single nodes are taken, and arranged into the pattern, putting nodes into full positions and skipping empty positions.
In the next step, the resulting structures are arranged into the same pattern.
All neighboring nodes in the resulting pattern are connected creating the fractal network. The fractal dimension of SC depends on the basic pattern.
Classical SC has $3\times 3$ pattern with all the squares full except the central one. Such SC has a fractal dimension $d=\frac{\ln 8}{\ln 3} \approx 1.893$.
We have investigated SC of dimensions ranging from $\frac{\ln 28}{\ln 8} \approx 1.6025$ to $\frac{\ln 8}{\ln 3}$.
Since it is impossible to numerically investigate true, infinite fractals, we will call the number of steps in what the network was made a level of fractal.
We have investigated SC with different patterns, that all have full squares along the pattern edges and empty interior, but differ in size. All SC fractals have {\em infinite} ramification.\\
The ramification is the minimal number of links that one has to remove to separate a part of any chosen size from a network.
The finite ramification means that the structure has some "weak points" where only a finite number of links connect together two parts of an infinite network. The infinite ramification means that infinite parts of infinite network are connected by infinite number of links.
For example, a regular square lattice has an infinite ramification, while a tree has a finite ramification order.\\

\begin{figure}
 \centerline{\epsfig{file=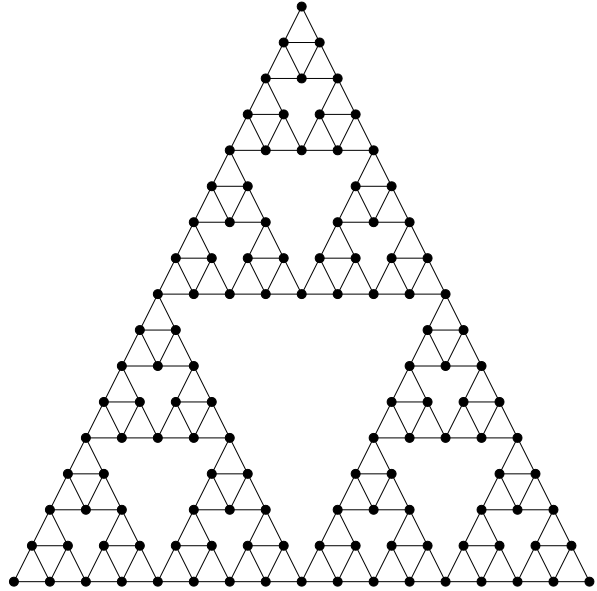,width=.6\columnwidth}}
    \caption{The construction of Sierpinski Gasket (SC) fractal network of level 5.}
    \label{gasket}
\end{figure}

The SG network is created in the following way (Fig.\ref{gasket}). Three nodes are taken and connected into a triangle.
In the middle of each edge a node is created and the three new nodes are connected between themselves.
This way the whole triangle is divided into four smaller ones.
In the next step all three non-central triangles are treated in the same way, adding nodes in middle of the triangle edges and linking them toghether.
SG has the fractal dimension $\frac{\ln 3}{\ln 2} \approx 1.5850$ and it possesses a {\em finite} ramification.\\
While in the case of SC, it was easy to create a general class of SC fractals with different fractal dimensions, we are not aware of any generalization of SC model that allows easy tuning of fractal dimensions.

\section{Results}
We have investigated ordering of the voter model in SC and SG fractals.
To measure the disorder, we have used the fraction of interfaces $\rho=I/E$,
where $I$ is a number of interfaces -- links connecting nodes with different spins, $E=N\mean{k}/2$ is the total number of links in the network.\\

\begin{figure}
 \centerline{\epsfig{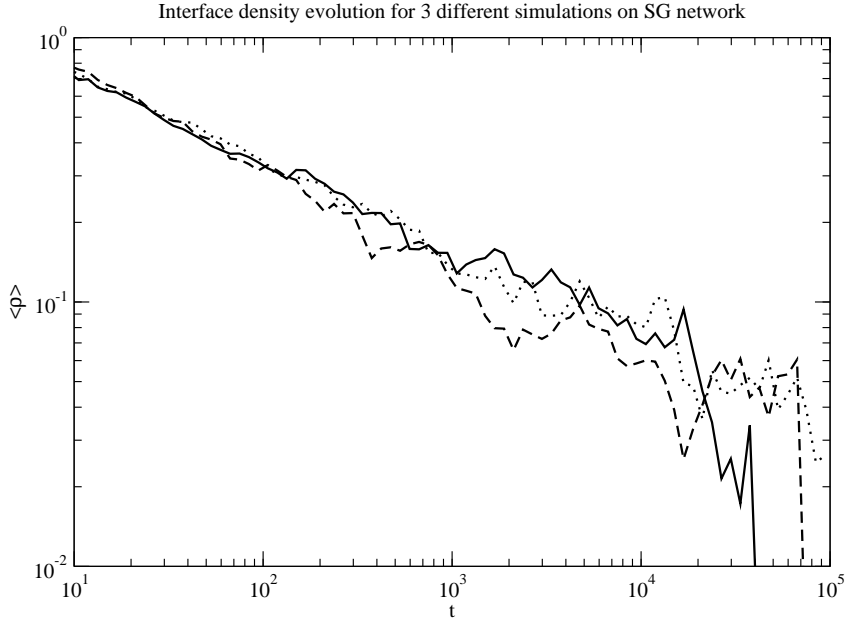}}
    \caption{The ordering process in the SG models for three different simulations. The data are for fractals of level 8.}
    \label{pojedyncze}
\end{figure}

The system orders (Fig.\ref{pojedyncze}) with the interface fraction $\rho$ decreasing as a power of time $t$.
However, due to the finite system size, there are fluctuations around the power-law.
Since the power-law decay becomes slower with time, the fluctuations become more significant, and they push the system into a completely ordered absorbing state after some time.\\

\begin{figure}
  \centerline{\epsfig{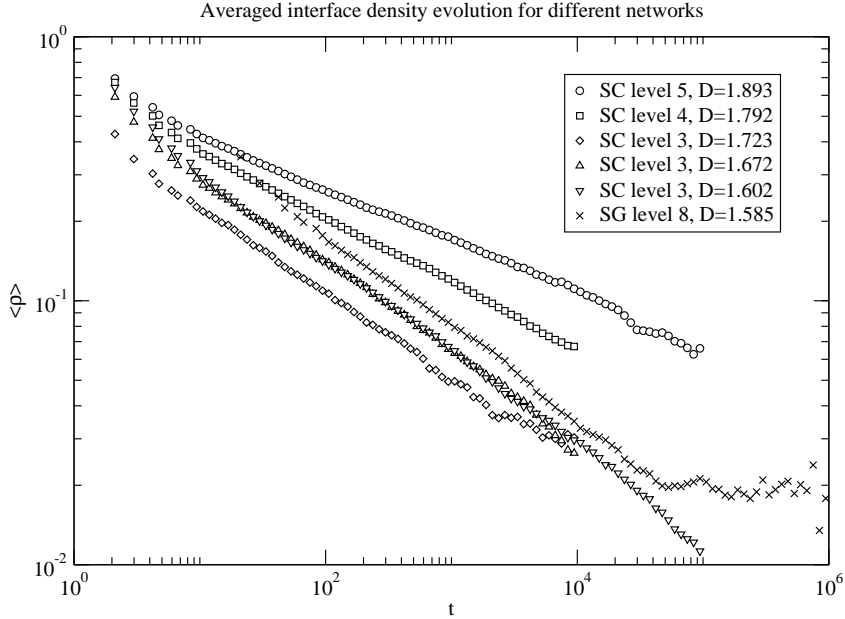}}
    \caption{The ordering process after averaging over simulations that did not order completely. The x marks are for SG, while other symbols are for SC. All the data are averaged over 100 network simulations and log-binned. The exponents $\alpha_{exp}$ are obtained from the slopes. The SG simulation data has been taken with 10 time steps intervals, thus the graph begins later than SC ones. We cannot explain the plateau for SG data, but in another simulations with higher level, but lower statistic, the plateau was absent.}
    \label{dynamika}
\end{figure}

To extract the power-law trend, we have averaged the results of many simulations, but to avoid the exponential decay due to complete ordering of the individual simulations, in a given time step we have averaged only over the simulations that were not completely ordered at that time. This way we have circumvented the fluctuations ordering the system and have obtained an approximation of an infinite network (Fig.\ref{dynamika}).\\
We observe the evolution of the interface fraction $\rho$ in time for networks with various fractal dimensions between $1$ and $2$. We have measured the exponent $\alpha$ of that power-law, and compare to the theoretical value \cite{krapivsky} (Eq.\ref{krap}).
\begin{table}
\label{tabelka}
\center{\begin{tabular} {ccccccc}
\hline type & symbol on fig.\ref{dynamika} & level & dimension & $\alpha_{theory}$ & $\alpha_{exp}$ & $\Delta \alpha_{exp}$\\
\hline
SC & circle & 5 & 1.8928 & 0.0536 & 0.1908 & 0.0007\\
SC & square & 4 & 1.7925 & 0.1038 & 0.2484 & 0.0007\\
SC & diamond & 3 & 1.7227 & 0.1387 & 0.3136 & 0.0029\\
SC & triangle up & 3 & 1.6720 & 0.1640 & 0.3362 & 0.0028\\
SC & triangle down & 3 & 1.6025 & 0.1988 & 0.3339 & 0.0017\\
SG & x mark & 9 & 1.5850 & 0.2075 & 0.3456 & 0.0034\\
regular & - & - & 1.0000 & 0.5000 & 0.4973 & 0.0006\\
\hline
\end{tabular}}
\caption{Theoretical and experimental exponents $\alpha$ (see Eq.\ref{krap}) for ordering processes in various networks. The results are averaged over 100 individual simulations. The levels of the fractals were maximized while keeping a number of nodes that allowed the actual simulations to be completed in reasonable amount of time. The regular network was a simple 1-dimensional chain with $\mean{k}=4$ and periodic boundary conditions.}
\vspace{.8cm}
\end{table}

\begin{figure}
 \centerline{\epsfig{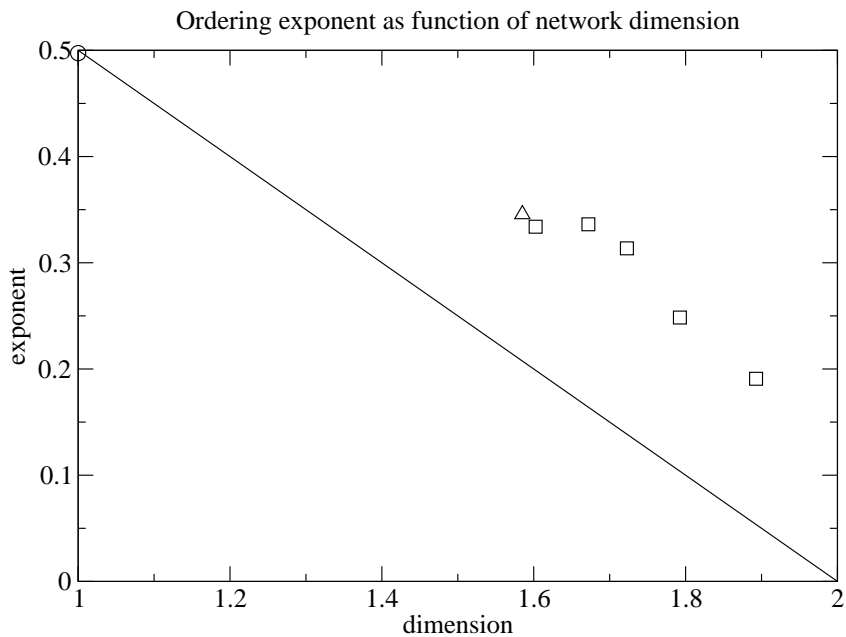}}
    \caption{Theoretical and experimental exponents for the ordering processes in various networks. The line is the analytic formula (Eq.\ref{krap}), the squares are exponents for SC networks, the triangle is exponent for SG network, the circle in the left upper corner is the exponent for the one-dimensional network. Error bars are no larger than symbol sizes.}
    \label{wyniki}
\end{figure}

\section{Conclusions}

The results (Fig.\ref{wyniki}) we have obtained show that the theoretical predictions \cite{krapivsky} (Eq.\ref{krap}) correctly describe the power-law ordering for a voter model on fractals,
but the exponents obtained through simulations differ significantly from those obtained from the analytic formula.\\
Moreover, the fact that the voter model in SC and SG behaves in the same fashion suggests that the ramification of the fractal does not influence the voter dynamics, unlike the Ising model case \cite{mandelbrot}.\\

This work was partially supported by a EU Grant {\it Measuring and Modelling Complex Networks Across Domains} (MMCOMNET) and by State Committee for Scientific Research in Poland (Grant No. 1P03B04727).


\begin{thebibliography}{7}

\bibitem{intro1} A. Aleksiejuk, J.A. Ho{\l}yst, D. Stauffer, Physica A 310 (2002), 260-266.
\bibitem{intro2} G. Bianconi, Phys. Lett. A 303 (2002), 166-168.
\bibitem{mandelbrot} Y. Gefen, B.B. Mandelbrot, A. Aharony, Phys. Rev. Lett. 45 (1980), 855.
\bibitem{ising1} J.M. Carmona, U.M.B. Marconi, J.J. Ruiz-Lorenzo, A. Taranc\'{o}n, Phys. Rev. B 58 (1998), 14387.
\bibitem{ising2} T. Sto\v{s}ic, B.D. Sto\v{s}ic, S. Milo\v{s}evi\'{c}, H.E. Stanley, Physica A 233 (1996), 31-38.
\bibitem{universality} I. Dornic, H. Chat\'{e}, J. Chave, H. Hinrichsen, Phys. Rev. Lett 87 (2001), 045701.
\bibitem{krapivsky} L. Frachenbourg, P.L. Krapivsky, Phys. Rev. E 53 (1996), 3009.

\end{thebibliography}
\end{document}